\documentclass[conference]{IEEEtran}
\author{Jun Tong  \\
 University of Wollongong, Wollongong, NSW 2522, Australia, Email: jtong@uow.edu.au 
}

\usepackage[normalem]{ulem} 
\usepackage{subcaption}
\usepackage[dvips]{graphicx}
\usepackage{amsmath,bm}
\usepackage{rotating}
\usepackage[strings]{underscore}
\usepackage{setspace}
\usepackage{color}
\usepackage{soul}
\usepackage{amsmath} 
\usepackage{amssymb}
\usepackage{cite}
\usepackage{stfloats}

\usepackage{tablefootnote}
\usepackage{algorithmicx}
\usepackage{algpseudocode}
\usepackage[normalem]{ulem}

\begin{document}

\title{On-Grid Equivalence of Continuous-Time Doubly Selective Channels: A Revisit of Bello's Models} 

\maketitle

\begin{abstract} 
  Significant studies on communications over doubly selective channels have utilized on-grid DD channel models, which are previously investigated in Bello's seminar paper in 1963. The DD grid is typically specified by the bandwidth and time duration of the transmission frames. However, the physical channels are determined by the propagation environments and they are typically off-grid.  Hence, there is often a gap between an actual physical channel and the on-grid model. This paper revisits the on-grid modeling of practical physical channels. We study the associated on-grid DD-domain representations for continuous-time, doubly selective channels with off-grid delay and Doppler shifts, accounting for practical time/frequency-domain windowing at the transceivers. The universal models obtained are applicable under the mild assumption that the windows have finite supports, and they extend Bello's classical results to account for more general windows. We also discuss the features and implications of the equivalent on-grid models. 
  
\end{abstract}

 \begin{IEEEkeywords}
    Delay-Doppler domain modulation, doubly selective channels, on-grid models   
 \end{IEEEkeywords}

\section{Introduction}
 Delay-Doppler (DD)-domain modulation including orthogonal time frequency space (OTFS)  and orthogonal DD  division multiplexing (ODDM) \cite{Hadani2017orthogonal, lin2022orthogonal, raviteja2018interference, Wei2021orthogonal, Gopalam2024zakotfs, Tong2023oddm} have attracted significant interests for communication and sensing over doubly selective (time-variant wideband) channels. Chirp-based waveforms such as orthogonal chirp division multiplexing (OCDM) \cite{ouyang2016orthogonal} and  affine frequency division multiplexing
(AFDM) \cite{bemani2021afdm} are also emerging as potential candidates for addressing the challenges of such doubly selective channels. They have demonstrated great potential for exploiting the stability and sparsity of the DD-domain channels and providing different tradeoff between performance and complexity.  

DD-domain channel models provide a basis for designing and analyzing the above modulation schemes. Many previous studies have assumed models with on-grid delay and Doppler shifts. This is natural and useful for examining the fundamental features of waveforms, such as their (bi)orthogonality with respect to the channel. Furthermore, on-grid models are well supported by Bello's seminar work on the modeling and analysis of doubly selective channels \cite{bello} in 1963, which shows that a general continuous-time linear time-variant (LTV) channel can be described using uniformly sampled delay and Doppler shifts, i.e., the channel can be expanded into multiple virtual, on-grid, continuous-time subchannels. This motivates the use of on-grid  channel models for waveform design and analysis.  

However, on-grid models should be used with careful accounting for the system configuration. For given physical channels and transceivers, the on-grid channel parameters are actually correlated. Furthermore, their cardinality and parameters can vary significantly with the system parameters and there can be necessary assumptions on the transceiver operations while applying the models. As such, it may not accurately reflect the actual system by using directly an on-grid model and presuming that it has the same sparsity and independence conditions as the underlying off-grid physical channels. Although statistical models of doubly selective channels  \cite{bello, matz2011fundamentals, hoeher1992statistical} have been highly valuable, deterministic on-grid models are also critical for accurate analysis and effective design of the DD-domain and chirp-based waveforms. This is especially relevant to integrated sensing and communication (ISAC) applications where precise information regarding the physical channel parameters needs to be gleaned from sampled observations using digital signal processing.   

In light of the above, we revisit the problem how a physical channel, in concatenation with transceiver windowing in the time or frequency domain, can be modeled using an equivalent DD channel with only on-grid delay and Doppler shifts, where the DD grid is specified by the window sizes. This problem was previously studied by Bello in 1963 \cite{bello} for rectangular windows. In this paper, we extend the results to consider more general windows with limited supports, which are deployed at the transmitter or receiver in the time or frequency domain. We show that despite the differences in the concrete implementations of the windows and instantaneous realizations of the channel, a unified on-grid model exists, which includes rectangular time- and frequency-domain windows applied at the transmitter or receiver, and subchannels with on-grid delay and Doppler shifts. In this model, the differences of the off-grid physical channels are captured by the correlated gains of the subchannels, whose expressions are explicitly given in this paper.  

 \begin{figure*}
\centering 
   \includegraphics[width=0.75\linewidth]{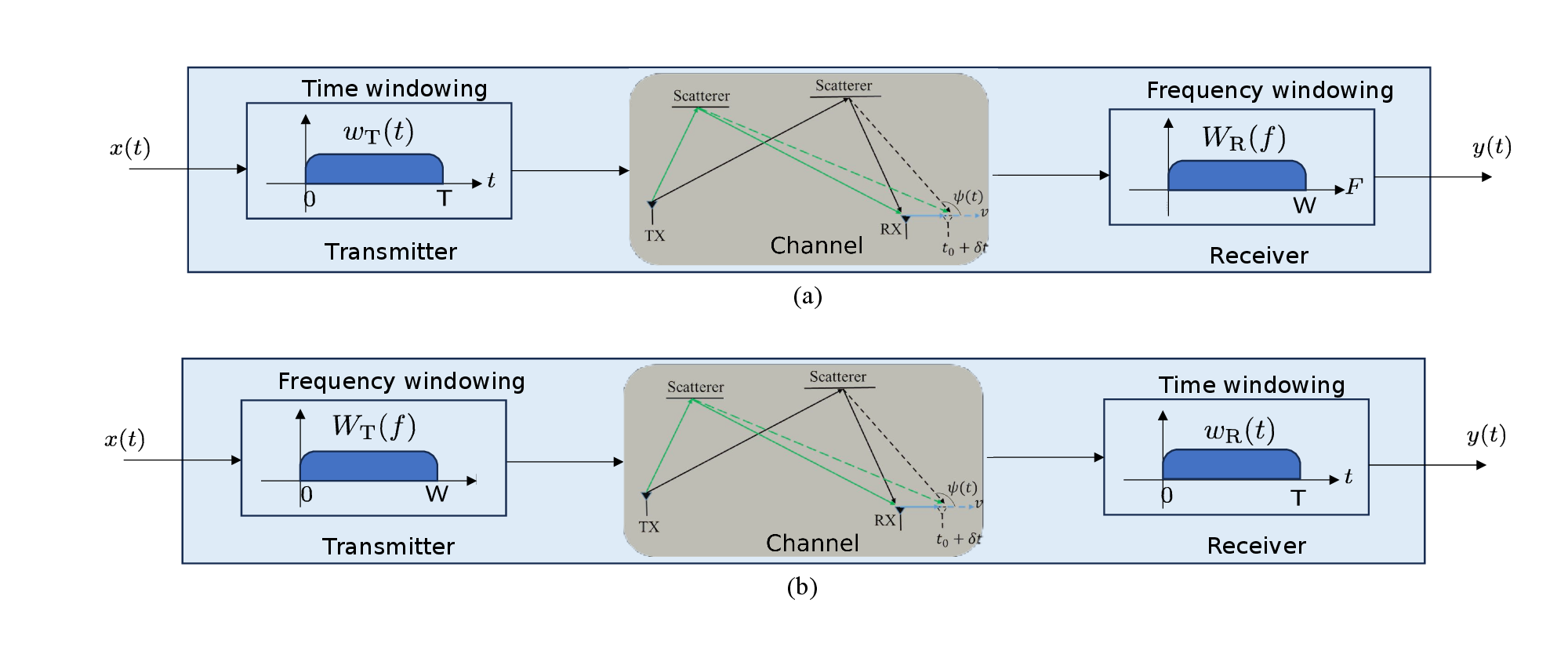}  
\caption{{DD channel with different transceiver constraints. (a): Time window at the transmitter and frequency window at the receiver; (b): Frequency window at the transmitter and time window at the receiver.}}
\label{fig:ddchannelconstrained}
\end{figure*} 

 \section{Spreading Functions for Doubly Selective Channels}

     Consider general LTV channels with continuously distributed scatterers, which cover discrete scatterers as special instances. Since we aim to focus on channel modeling in this paper, the noise will be ignored but we note that the receiver operations employed may affect their statistics. We first review several useful channel spreading functions following \cite{bello}. The {\emph{input delay-spread function}} $h(t, \tau)$ produces the channel output signal $y(t)$ by first delaying the input signal $x(t)$ and then amplifying as 
      \begin{equation} 
      \label{ythttau}
      \begin{aligned}
             y(t) = \int x(t-\tau) h(t, \tau) d \tau.       
      \end{aligned}
      \end{equation}  
    Alternatively, the { \emph{output delay-spread function}} $q(t,\tau)$ can produce the same output signal by first amplifying the input signal and then delaying as  
    \begin{equation} 
    \label{ytqtaut}
    \begin{aligned}
         y(t) = \int x(t-\tau) q(t-\tau, \tau) d \tau.       
    \end{aligned}
    \end{equation} 
    Clearly, for the same physical channel, the two models $h(t, \tau)$ and $q(t,\tau)$ are related. For example, for a single-scatterer channel with delay shift $\widetilde \tau_p$, Doppler shift $\widetilde \nu_p$ and path gain $\widetilde \rho_p$, if we have $h(t, \tau) = \widetilde{\rho}_p e^{j 2 \pi \widetilde{\nu}_p (t-\widetilde{\tau}_p) } \delta(\tau-\widetilde{\tau}_p)$, where $\delta(\cdot)$ denotes the Dirac delta function, then $q(t, \tau) = \widetilde{\rho}_p   e^{j 2 \pi \widetilde{\nu}_p t } \delta(\tau-\widetilde{\tau}_p)$. Clearly, both $h(t,\tau)$ and $q(t,\tau)$ link the transmitted and received signals in the time domain, and they are related by $q(t,\tau)=h(t+\tau, \tau)$. Further explanations can be found in \cite{bello}. 

    The IO relation of the channel can be given by using alternative combinations of the time- and frequency-domain representation of the input and output signal. Using the { \emph{time-variant transfer function}}   
      \begin{equation} 
       \label{Tft}
       \begin{aligned}
             T(f,t) = \int h(t,\tau) e^{-j2\pi f \tau} d \tau,         
      \end{aligned}
      \end{equation}  
    we have 
    \begin{equation} 
    \label{ytXfTft}
    \begin{aligned}
         y(t) = \int  T(f,t) X(f) e^{j2\pi f t} d f,       
    \end{aligned}
    \end{equation}   
    where $X(f)$ represents the input signal in the frequency domain. Similarly, we have 
    \begin{equation} 
    \label{yfxtTft}
    \begin{aligned}
         Y(f) = \int x(t) M(t, f) e^{-j2\pi f t} d t ,       
    \end{aligned}
    \end{equation} 
    where the { \emph{frequency-dependent modulation function}} is given by 
    \begin{equation} 
    \label{Mtf}
    \begin{aligned}
         M(t,f) = \int q(t,\tau) e^{-j2\pi f \tau} d \tau.         
    \end{aligned}
    \end{equation} 
  
  Other spreading functions include the { \emph{delay-Doppler spread function}} $ U(\tau,\nu)$ defined as  
  \begin{equation} 
  \label{Utaunu}
  \begin{aligned}
       U(\tau,\nu)   & = \int  h(t,\tau) e^{- j2\pi \nu t} d t,            
  \end{aligned}  
  \end{equation} 
  which models the output signal 
  \begin{equation} 
  \label{ytUtaunu}
  \begin{aligned}
       y(t) & = \int \int  U(\tau,\nu)  x(t-\tau) e^{j 2 \pi \nu t}  d \tau d\nu         
  \end{aligned}  
  \end{equation} 
  by first delaying the input signal and then modulating (Doppler shifting). From (\ref{Tft}) and (\ref{Utaunu}), we further have 
  \begin{equation}
      T(f,t) = \int \int  U(\tau,\nu) e^{ j2\pi \nu t} e^{- j2\pi \tau f} d \nu d \tau.  
  \end{equation}
  
  Alternatively, the  { \emph{Doppler-delay spread function}} is defined by 
  \begin{equation} 
  \label{Vnutauwrtqttau}
  \begin{aligned}
        V(\nu,\tau) =  \int  q(t,\tau) e^{- j2\pi \nu t} d t.              
  \end{aligned}
  \end{equation} 
    It is widely used in the literature on DD-domain and chirp-based modulation for describing the time-domain IO relation of DD-domain modulation as 
  \begin{equation} 
  \label{Vnutautoyt}
  \begin{aligned}
           y(t) = \int \int V(\nu,\tau) x(t-\tau) e^{j 2 \pi \nu (t-\tau)}  d \tau d\nu,              
  \end{aligned}
  \end{equation} 
 where the input signal is first Doppler-shifted and then delayed.
 In the frequency domain, the input and output are related as  
   \begin{equation} 
  \label{Vnutautoyt}
  \begin{aligned}
           Y(f) = \int \int V(\nu,\tau) X(f-\nu) e^{-j 2 \pi f \tau }     d \tau d\nu.              
  \end{aligned}
  \end{equation}

 It should be noted that all the above channel spreading functions originate from the same physical channel and they are related via Fourier transforms or scaling. For example,   
  \begin{equation} 
  \label{VUrelation}
  \begin{aligned}
        V(\nu,\tau) =  U(\tau,\nu) e^{ j2\pi \nu \tau},              
  \end{aligned}
  \end{equation} 
  \begin{equation} 
  \label{htauttv}
  \begin{aligned}
         h(t, \tau) & = \int_{\nu} V(\nu, \tau)   e^{j 2 \pi \nu   {(t-\tau)} } d \nu,        
  \end{aligned}
  \end{equation}   
 and from (\ref{Mtf}) and (\ref{Vnutauwrtqttau}), we have  
  \begin{equation}
  \begin{aligned}
    M(t,f) & = \int \int V(\nu, \tau) e^{j 2 \pi \nu t}  e^{-j2 \pi \tau f} d\tau d \nu. 
  \end{aligned}
  \end{equation}  
 The above channel spreading functions exhibit different speeds of variation over time. For example, the input delay-spread function $h(t,\tau)$ and the Doppler-delay spread function $V(\nu, \tau)$ can be regarded as approximately constant over a coherence time and a stationary time, respectively, with the latter being significantly longer in many applications. This may be one of the motivations for developing DD-domain modulation. When the support of the delay-Doppler and Doppler-delay spread functions $U(\tau,\nu)$ and $V(\nu,\tau)$ is limited to the regions with low delay and Doppler in the DD plane, the channel is regarded as \emph{underspread}, which is assumed by many studies on DD-domain modulation.

 \begin{figure*}
    \centering
  \subfloat[ {Input time-windowed, output frequency-windowed}]{
       \includegraphics[width=0.8 \linewidth]{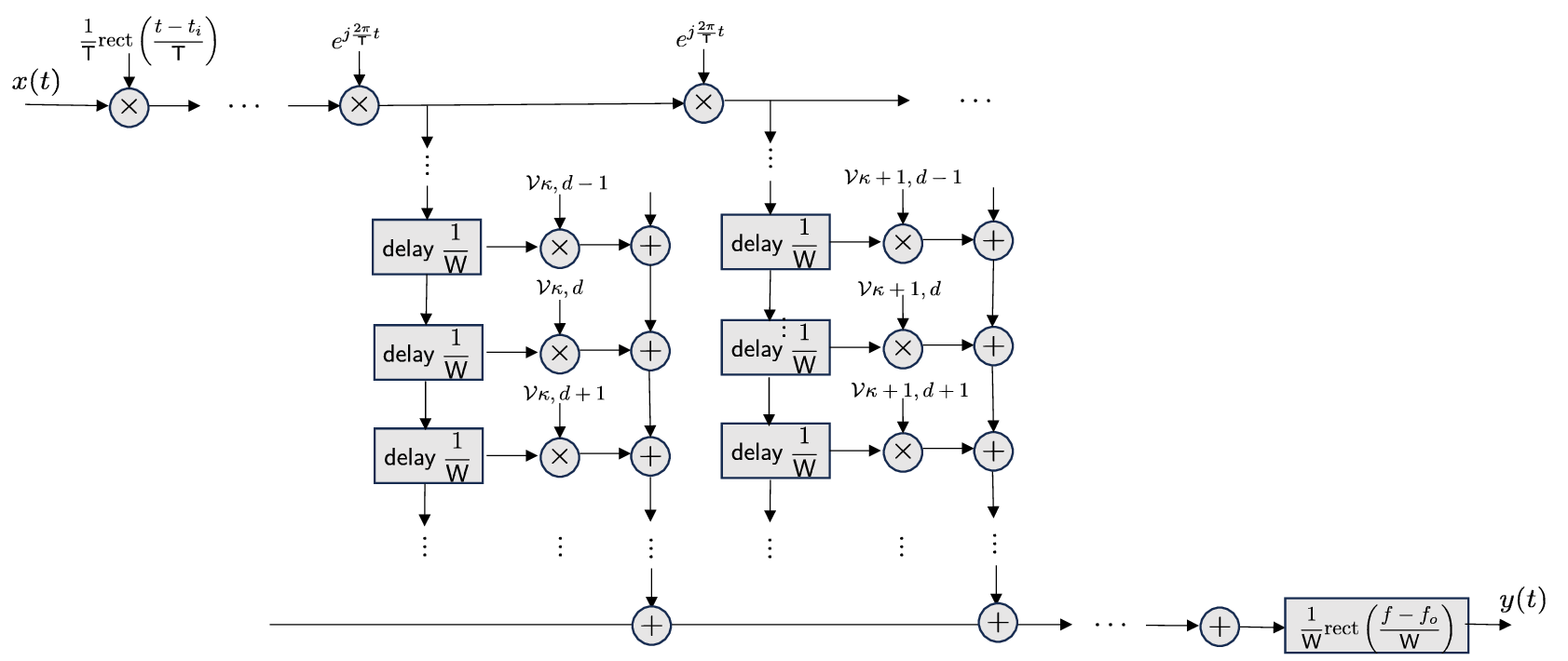}}  \\ 
  \subfloat[ {Input frequency-windowed, output time-windowed}]{%
        \includegraphics[width=0.8 \linewidth]{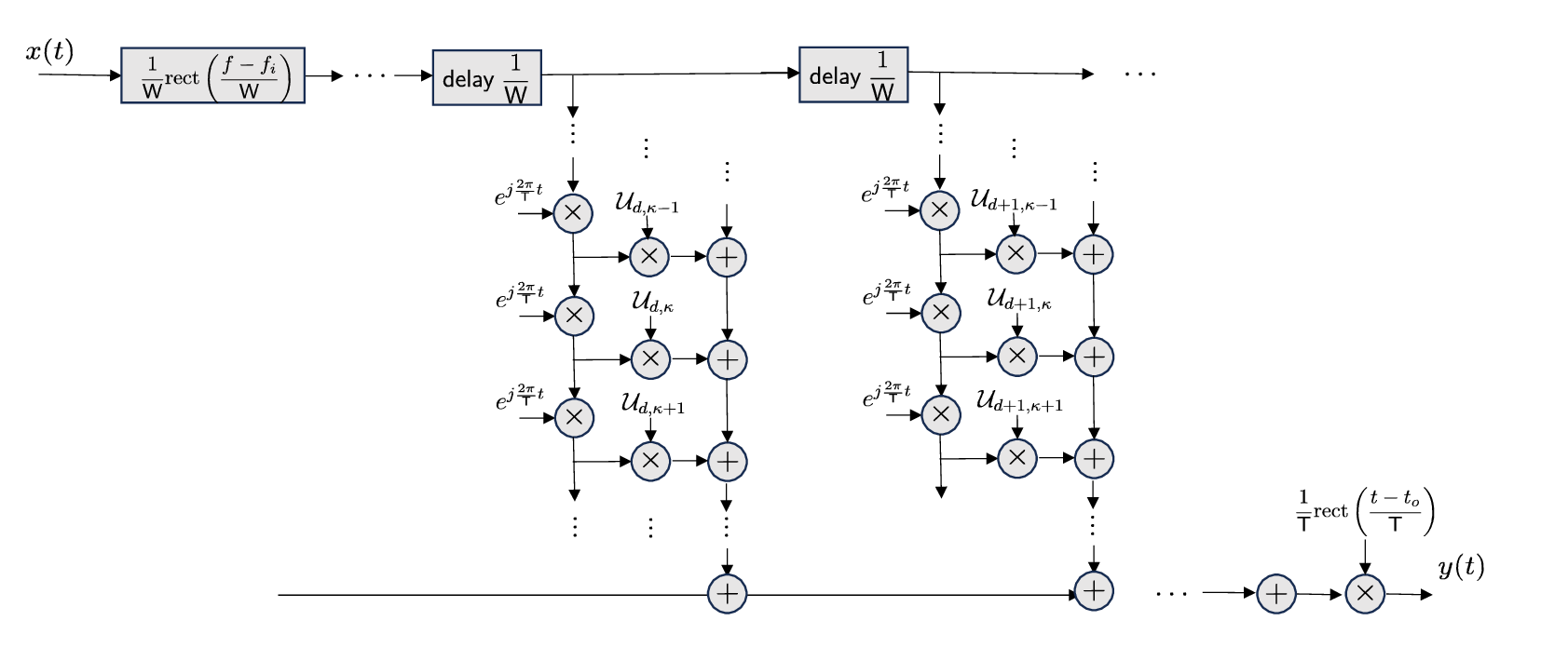}}             
    \caption{Equivalent DD channel models for systems with time-frequency windowing in Fig. 1(a) and (b).} 
  \label{FigESDDChannels} 
\end{figure*}

\section{Continuous-Time On-Grid DD Channel Models Subject to Time-Frequency (TF) Windowing}
\label{sec::ongridcontinuous}

Practical communication systems operate with finite time-frequency (TF) resources, which can be regulated by using windows in the time or frequency domain. Given the frame duration and bandwidth of transmission, an observed DD channel may have limited degrees of freedom and can be characterized using canonical models \cite{bello} with discrete DD shifts. With such models, the IO model describes the output signal as a superposition of delay- and Doppler-shifted versions of the input signal, with uniformly sampled delay and Doppler shifts. This helps demonstrate the coupling of the waveform and the channel, and enables the derivation of signal processing solutions. Following \cite{bello}, this paper derives such models by exploiting the sampling theorem and the Whittaker–Shannon interpolation formula. One mild assumption under the on-grid (sampled) model is that the effective channel (including the concatenation with the transceiver windows) has finite time or frequency supports.  

We consider a time or frequency constraint imposed on the input or output of the channel, respectively, in light that no signal can be simultaneously limited in time and frequency. We consider the two cases in Fig. \ref{fig:ddchannelconstrained} (a) and (b) and establish the on-grid equivalent channel models. We will use $w_{\mathrm{T}}(t)$, $W_{\mathrm{R}}(f)$, etc., to denote the window function at the transmitter and receiver, in the time or frequency domain, respectively. The study may be extended to other cases of interests.  

\subsection{Transmitter Time Windowing and Receiver Frequency Windowing} 

Within this category, where the transmitted and received signal is subject to a time and frequency window, respectively, the special case with rectangular windows has been discussed in \cite{bello}. We here extend the discussion to more general windows. In this case, the IO relation of (\ref{yfxtTft}) is revised as 
\begin{equation} 
\label{yfxtTftwindowed}
\begin{aligned}
Y(f) & = \int x(t) w_{\mathrm{T}}(t) M(t, f) W_{\mathrm{R}}(f)  e^{-j2\pi f t} d t,       
\end{aligned} 
\end{equation}
where the windows have finite supports centered at $t_i$ and $f_o$ with durations of $\mathsf{T}$ and $\mathsf{W}$ due to the input time constraint and output bandwidth constraint:  
\begin{equation}
    \begin{aligned}
         w_{\mathrm{T}}(t): & t_i - \frac{\mathsf{T}}{2}  \le t \le t_i +  \frac{\mathsf{T}}{2}, \\ 
         W_{\mathrm{R}}(f): & f_o -\frac{\mathsf{W}}{2}\le f \le f_o+ \frac{\mathsf{W}}{2}.
    \end{aligned}
\end{equation} 
Clearly, the effective channel including the transceiver windowing can be described by the effective \emph{frequency-dependent modulation function} as 
\begin{equation} 
\label{Mtfwindowed}
\begin{aligned}
M'(t,f) =  w_{\mathrm{T}}(t) M(t, f) W_{\mathrm{R}}(f).           
\end{aligned} 
\end{equation}
The corresponding effective \emph{Doppler-delay spread function} can then be found as 
\begin{equation} 
\label{VnutauTFwindowed}
\begin{aligned}
V'(\nu, \tau) & = \int \int M'(t,f) e^{-j 2\pi \nu t} e^{j 2\pi \tau f} d f d t.   
\end{aligned} 
\end{equation} 
Because $M'(t,f)$ is limited in $t$, given $f$, its Fourier transform w.r.t. $t$  can be represented by its samples in combination with interpolation. Similar treatments are available for $f$ and $\tau$. The resulting representations can be derived by examining a single scatterer as the overall channel is linear and the superposition principle applies. 

Now, let us consider an arbitrary single-path physical channel with \emph{Doppler-delay spread function} $V(\nu, \tau)=\widetilde{\rho}_p \delta(\nu-\widetilde{\nu}_p)\delta(\tau-\widetilde{\tau}_p)$ and \emph{frequency-dependent modulation function}   
\begin{equation}
\begin{aligned}
    M(t,f) & = \int \int V(\nu, \tau) e^{j 2 \pi \nu t}  e^{-j2 \pi \tau f} d\tau d \nu  \\
           & = \widetilde{\rho}_p e^{j 2 \pi \widetilde{\nu}_p t} e^{-j 2 \pi \widetilde{\tau}_p f}. 
\end{aligned}
\end{equation}
Then, from (\ref{VnutauTFwindowed}) we have 
{\small \begin{equation} 
\label{VnutauTFwindowedv2}
\begin{aligned}
V'(\nu, \tau) 
 & = \widetilde{\rho}_p   \int w_{\mathrm{T}}(t)  e^{j 2 \pi \widetilde{\nu}_p t}  e^{-j 2\pi \nu t}   d t      \int  W_{\mathrm{R}}(f) e^{-j 2 \pi \widetilde{\tau}_p f}    e^{j 2\pi \tau f} d f   \\ 
 & = \widetilde{\rho}_p W_{\mathrm{T}} (\nu - \widetilde{\nu}_p)  w_{\mathrm{R}} (\tau - \widetilde{\tau}_p)   
\end{aligned} 
\end{equation}}    
where we have the Fourier transform (FT) and inverse FT (IFT) relationships:  
\begin{equation}
    \begin{aligned}
        W_{\mathrm{T}} (\nu  ) & \triangleq  \int w_{\mathrm{T}}(t)    e^{-j 2\pi \nu t}   d t,   \\
        w_{\mathrm{R}} (\tau ) & \triangleq  \int W_{\mathrm{R}}(f)    e^{ j 2\pi \tau f}   d f.  
    \end{aligned}
\end{equation}

Due to the time and frequency constraints, the sampling theorem suggests that $V'(\nu, \tau)$ can be expanded using its discrete samples, where the sampling interval is determined by the bandwidth and time duration constraints  $\mathsf{W}$  and $\mathsf{T}$. 
By noting that the original windows can be viewed as passband signals, we can represent the effective \emph{Doppler-delay spread function} as (\ref{Vprimenutau}) on the top of next page,  
\begin{figure*}
\begin{equation} 
\label{Vprimenutau}
\begin{aligned}
V'(\nu, \tau) & = \widetilde{\rho}_p \sum_{\kappa=-\infty}^{\infty} \sum_{d=-\infty}^{\infty}   e^{j 2\pi f_o (\tau-\tau_d)}   w_{\mathrm{R}} (\tau_d-\widetilde{\tau}_p)  \frac{\sin(\pi \mathsf{W} (\tau-\tau_d))}{\mathsf{W} (\tau-\tau_d)} e^{-j 2\pi t_i (\nu-\nu_\kappa) }    W_{\mathrm{T}} (\nu_\kappa - \widetilde{\nu}_p)  \frac{\sin(\pi \mathsf{T} (\nu-\nu_\kappa))}{\mathsf{T} (\nu-\nu_\kappa)}  \\ 
              & =  \sum_{\kappa=-\infty}^{\infty} \sum_{d=-\infty}^{\infty} \mathcal{V}_{\kappa, d}  e^{j 2\pi f_o (\tau-\tau_d)}   \frac{\sin(\pi\mathsf{W} (\tau-\tau_d))}{\mathsf{W}(\tau-\tau_d)} e^{-j 2\pi t_i (\nu-\nu_\kappa) }   \frac{\sin(\pi\mathsf{T}  (\nu-\nu_\kappa))}{\mathsf{T} (\nu-\nu_\kappa)} \\
{\mathcal{V}}_{\kappa, d} &  = \widetilde{\rho}_p      W_{\mathrm{T}} (\nu_\kappa - \widetilde{\nu}_p)  w_{\mathrm{R}} (\tau_d-\widetilde{\tau}_p)            
\end{aligned} 
\end{equation}    
\end{figure*}
where 
$\nu_\kappa \triangleq \frac{\kappa}{\mathsf{T}},  \tau_d \triangleq \frac{d}{\mathsf{W}}$        
are the sampled Doppler and delay shifts, uniformly located on the DD grid specified by the duration of the time window and bandwidth of the frequency window. 
From (\ref{Vprimenutau}) we obtain (\ref{Mtfequivalentbutusesamples}), 
\begin{figure*}
\begin{equation} 
\label{Mtfequivalentbutusesamples}
\begin{aligned}
M'(t,f) & = \int \int V'(\nu, \tau) e^{j2\pi \nu t}  e^{-j2\pi  \tau f} d\tau d \nu \\
        & =  \sum_{\kappa=-\infty}^{\infty} \sum_{d=-\infty}^{\infty} \widetilde{\mathcal{V}}_{\kappa, d}  
        \int e^{j 2\pi f_o (\tau-\tau_d)}   \frac{\sin(\pi\mathsf{W} (\tau-\tau_d))}{\mathsf{W}(\tau-\tau_d)}  e^{-j2\pi  \tau f} d\tau 
        \int e^{-j 2\pi t_i (\nu-\nu_\kappa) }   \frac{\sin(\pi\mathsf{T}  (\nu-\nu_\kappa))}{\mathsf{T} (\nu-\nu_\kappa)}  e^{j2\pi \nu t} d \nu   \\
        & = \frac{1}{\mathsf{W}} \mathrm{rect}\left( \frac{f-f_o}{\mathsf{W}} \right) \left(  \sum_{\kappa=-\infty}^{\infty} \sum_{d=-\infty}^{\infty} {\mathcal{V}}_{\kappa, d}  e^{-j2\pi  \tau_d f} e^{ j2\pi  \nu_\kappa t} \right)    \frac{1}{\mathsf{T}} \mathrm{rect}\left( \frac{t-t_i}{\mathsf{T}} \right)  
\end{aligned} 
\end{equation}    
\end{figure*}
which shows that the effective channel consists of three elementary functional blocks: A \emph{rectangular time window} at the transmitter; a composite DD channel consisting of subchannels formed by on-grid virtual scatterers; and a \emph{rectangular frequency window} at the receiver. This is illustrated in Fig. \ref{FigESDDChannels}(a). The equivalent \emph{frequency-dependent modulation function}  $\mathcal{V}_{\kappa, d}  e^{-j2\pi f \tau_d} e^{ j2\pi t \nu_\kappa }$ corresponds exactly to a \emph{Doppler-delay spread function} $ {\mathcal{V}}_{\kappa, d} \delta(\tau-\tau_d) \delta(\nu-\nu_\kappa)$.    
Therefore, the composite DD channel has an effective Doppler-delay spread function 
\begin{equation}
    \widetilde{V}(\nu, \tau) = \sum_{\kappa=-\infty}^{\infty} \sum_{d=-\infty}^{\infty}   \mathcal{V}_{\kappa, d} \delta(\nu-\nu_\kappa) \delta(\tau-\tau_d),  
\end{equation}
where $\mathcal{V}_{\kappa, d}$ are given by (\ref{Vprimenutau}) for the case with a single physical path. 
 For more general physical channels with continuous scatterers, we have 
\begin{equation} 
\label{Vmncontinuous}
\begin{aligned}
 \mathcal{V}_{\kappa,d}   = \int \int V(\nu, \tau) W_{\mathrm{T}} (\nu_\kappa - \nu )   w_{\mathrm{R}} (\tau_d-\tau) d\nu d\tau,              
\end{aligned} 
\end{equation}    
which can be further specified to cases with $P$ discrete physical scatterers as 
\begin{equation} 
\label{Vmndiscretea}
\begin{aligned}
     \mathcal{V}_{\kappa, d}   =\sum_{p=1}^P V(\widetilde{\nu}_p, \widetilde{\tau}_p) W_{\mathrm{T}} (\nu_\kappa - \widetilde{\nu}_p)  w_{\mathrm{R}} (\tau_d-\widetilde{\tau}_p).                  
\end{aligned} 
\end{equation}    

From the above, a physical DD channel together with TF windows at the transceiver can be represented using on-grid DD channels. The model also reveals the potential spread of a physical channel path onto multiple DD channel taps, according to the shapes of the window functions, as shown in (\ref{Vmndiscretea}). Even a single physical propagation path can result in multiple on-grid DD channel taps, which can lead to inter-symbol-interference (ISI). 
Clearly, the gains $\mathcal{V}_{\kappa, d}$ associated with the virtual scatterers are correlated, which may be considered when simulating/modeling DD channels using their on-grid equivalence to capture the mechanism underlying their generation. In the special case of rectangular windows, we have 
\begin{equation}
\label{rectwindows}
\begin{aligned}
W_{\mathrm{T}} (\nu ) & \triangleq \mathsf{T}   e^{-j 2 \pi  t_i \nu}  \frac{\mathrm{sin} (\pi \mathsf{T}  \nu)}{\pi \mathsf{T}  \nu} = \mathsf{T}   e^{-j 2 \pi  t_i \nu}  \mathrm{sinc} ( \mathsf{T}  \nu),  \\ 
w_{\mathrm{R}} (\tau) & \triangleq \mathsf{W}  e^{ j 2 \pi f_o \tau}  \frac{\mathrm{sin} (\pi \mathsf{W}  \tau)}{\pi \mathsf{W}  \tau}  = \mathsf{W}  e^{ j 2 \pi f_o \tau}   \mathrm{sinc} (  \mathsf{W}  \tau),       
\end{aligned}
\end{equation}
and 
\begin{equation}
\label{Vkappad}
\begin{aligned}
 \mathcal{V}_{\kappa, d} & =\mathsf{T} \mathsf{W}  \int \int  V(\nu, \tau)  e^{-j 2 \pi t_i  (\nu_\kappa -\nu)   }  e^{ j 2 \pi f_o (\tau_d -\tau )   }   \times \\
                 & \qquad \mathrm{sinc} \left( \mathsf{T} \left(\nu_\kappa - \nu \right)  \right) \mathrm{sinc} \left( \mathsf{W} \left( \tau_d - \tau \right) \right) d \nu d \tau.    
\end{aligned}
\end{equation}  
This agrees with Bello's model given by \cite[Equation (3)]{bello1969measurement}. Note that there are long tails in the transformed domains for rectangular windows. Thus, it is possible that a large number of on-grid subchannels can arise from a single physical path.

\subsection{Transmitter Frequency Windowing and Receiver Time Windowing} When a frequency window $W_{\mathrm{T}}(f)$ centered at $f_i$ is applied at the transmitter and a time window $w_{\mathrm{R}}(t)$ centered at $t_o$ is applied at the receiver, similar equivalent on-grid models can be derived from the windowed transfer function 
\begin{equation} 
\label{Tftwindowed}
\begin{aligned} 
    T'(f,t) =  w_{\mathrm{R}}(t) T(f, t) W_{\mathrm{T}}(f).           
\end{aligned} 
\end{equation} 
The effective channel can be modeled as (\ref{Tftequivalentbutusesamples}) on the top of this page and illustrated in Fig. \ref{FigESDDChannels}(b), where    
\begin{figure*}
\begin{equation} 
\label{Tftequivalentbutusesamples}
\begin{aligned}
      T'(f,t)  
         = \frac{1}{\mathsf{T}} \mathrm{rect}\left( \frac{t-t_o}{\mathsf{T}} \right) \left(  \sum_{\kappa=-\infty}^{\infty} \sum_{d=-\infty}^{\infty} \mathcal{U}_{d,\kappa}  e^{-j2\pi f \tau_d} e^{ j2\pi t \nu_\kappa } \right)    \frac{1}{\mathsf{W}} \mathrm{rect}\left( \frac{f-f_i}{\mathsf{W}} \right)  
\end{aligned} 
\end{equation}    
\end{figure*} 
\begin{equation} 
\label{Umncontinuous}
\begin{aligned}
     \mathcal{U}_{d,\kappa}   = \int \int U(\tau, \nu)   w_{\mathrm{T}} (\tau_d-\tau) W_{\mathrm{R}} (\nu_\kappa - \nu ) d\nu d\tau             
\end{aligned} 
\end{equation}   
for continuous scatterers and 
\begin{equation} 
\label{Umndiscrete}
\begin{aligned}
    \mathcal{U}_{d,\kappa}   =\sum_{p=1}^P U( \widetilde{\tau}_p, \widetilde{\nu}_p)    w_{\mathrm{T}} (\tau_d-\widetilde{\tau}_p) W_{\mathrm{R}} (\nu_\kappa - \widetilde{\nu}_p)                
\end{aligned} 
\end{equation}    
for discrete scatterers, 
  with $w_{\mathrm{T}} (\tau)$ and $W_{\mathrm{R}} (\nu)$ being respectively the IFT and FT of the windows. 
The above modeling confirms that we can use rectangular windows together with appropriate on-grid DD channels to model a general channel with general TF windows at the transceiver. 
The input-output relation can  be rewritten as 
\begin{equation}
\label{basebandmodelDelayDoppler}
  y(t) = \sum_{\kappa} \sum_{d}  \mathcal{U}_{d,\kappa}  
  x\left(t-\tau_d\right) e^{j 2 \pi \nu_\kappa t}.     
\end{equation}  

It should be noted that in the on-grid models in Fig. \ref{FigESDDChannels}, in theory, there can be an infinite number of on-grid subchannels described by $\{\mathcal{V}_{\kappa, d}\}$ or $\{\mathcal{U}_{d,\kappa}\}$. In practice, the cardinality and gains of the on-grid model depend on the window shape and the instantaneous realization of the physical channel.

\section{Numerical Examples}
    We next demonstrate the above ESDD channel models established for continuous-time doubly selective channels. We consider a system with transmission duration $\mathsf{T}= \frac{16384}{\mathsf{W}}$ 
    and receiver bandwidth $\mathsf{W}=7.68$ MHz. The time-domain rectangular and Hamming window with duration $\mathsf T$ are considered for the transmitter, while the raised cosine (RC) filter with bandwith $\mathsf W$ but different rolloffs is considered as the frequency-domain window applied at the receiver.  
    
\begin{figure}
    \centering 
       \includegraphics[width=1 \linewidth]{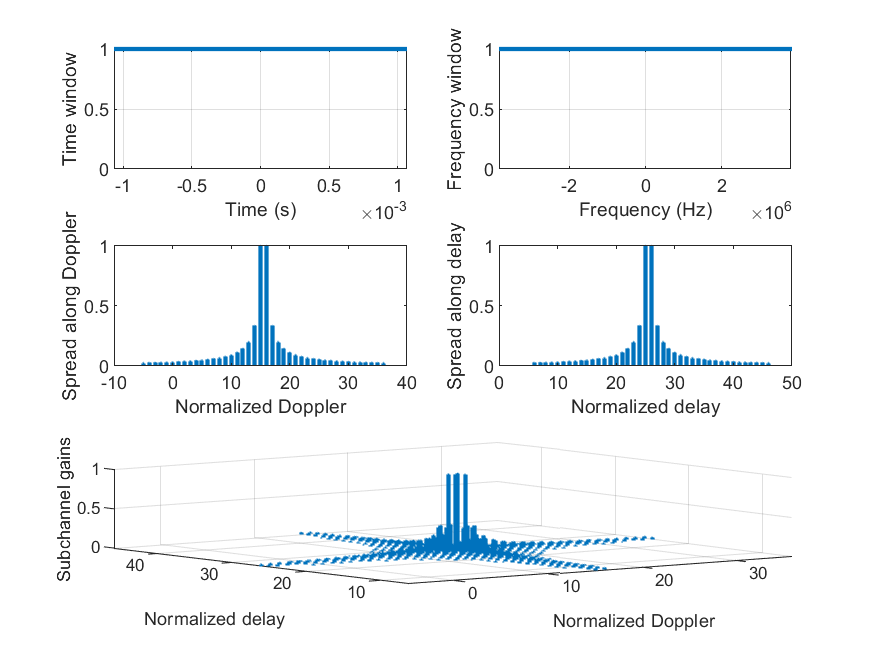}  
    \caption{\small Magnitudes of the spread of an off-grid physical channel onto the on-grid equivalence with rectangular windows applied.}
    \label{fig:chapter02:continuousesddrectangularwindows}
    \end{figure}

\begin{figure}
    \centering 
       \includegraphics[width=1 \linewidth]{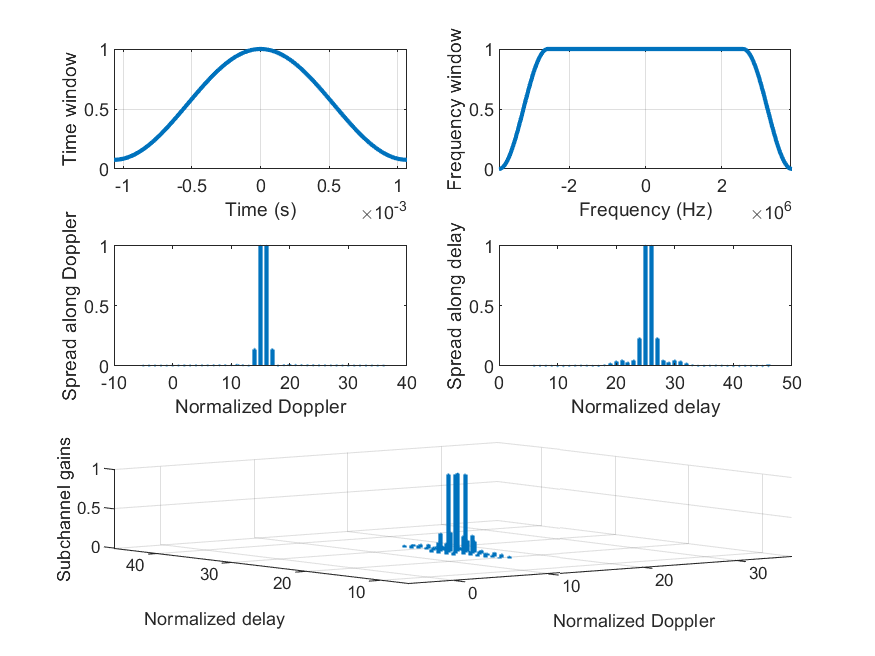}  
    \caption{\small Magnitudes of the spread of an off-grid physical channel onto the on-grid equivalence with smoother windows, where the time-domain Hamming window is applied at the transmitter and the frequency-domain RC window with rolloff factor $0.2$ is applied at the receiver.}
    \label{fig:chapter02:continuousesddsmootherwindows}
    \end{figure}

    Fig. \ref{fig:chapter02:continuousesddrectangularwindows} and \ref{fig:chapter02:continuousesddsmootherwindows} show numerical examples of $\{\mathcal{V}_{\kappa, d}\}$ for a physical channel with a single off-grid path with parameters $\widetilde{\tau}_p= \frac{25.5}{\mathsf{W}}$ and $\widetilde{\nu}_p = \frac{15.5}{\mathsf{T}}$, where the window shapes, the spread of the channel along delay, Doppler, and two-dimensional DD grid are demonstrated. Results are normalized for ease of visualization. From (\ref{Vprimenutau}), given a single-path physical channel, the spreads of the path along delay and Doppler are decoupled, and determined by the physical path parameters and the (inverse) Fourier transforms of the frequency- or time-domain windows, respectively. It is seen that smoother windows (e.g., the time-domain Hamming window and the raised cosine filter) result in less spread of the channel on the DD grid compared to that with rectangualr windows.

\begin{figure}
    \centering 
       \includegraphics[width=1.0 \linewidth]{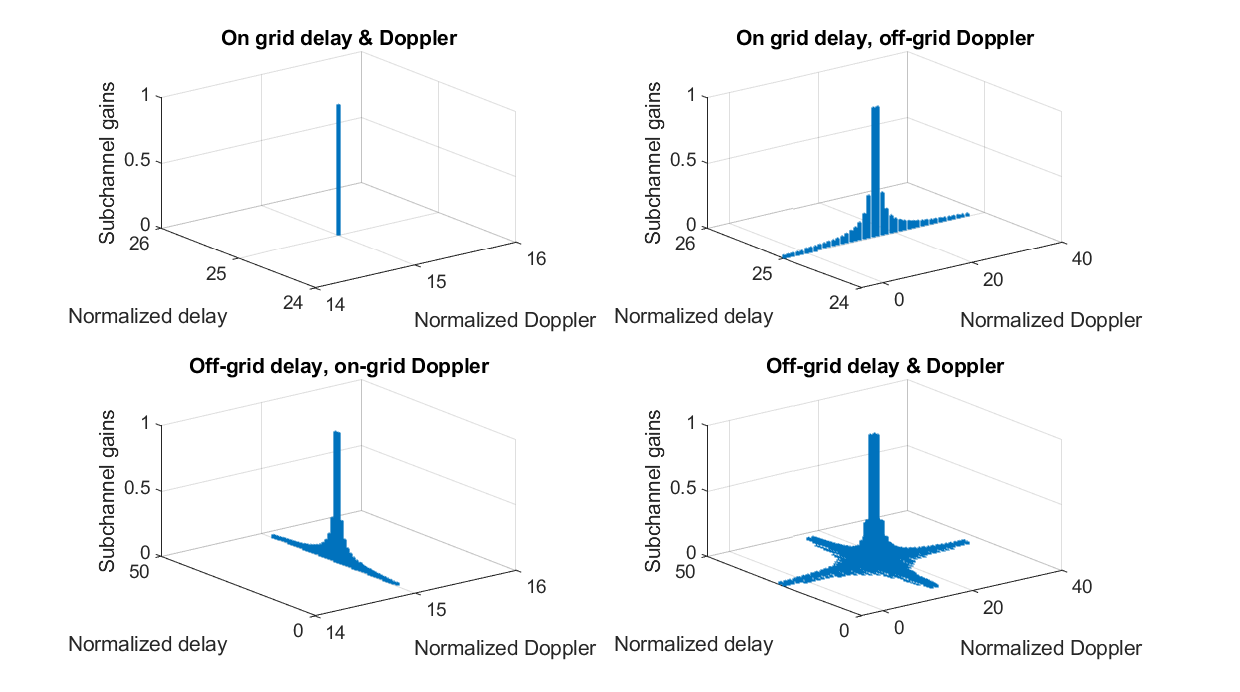}  
    \caption{\small Magnitudes of the spread of an off-grid physical channel path onto the on-grid equivalence with rectangular windows, different cases of the physical path parameters are considered.}
    \label{fig:chapter02:continuousesddrectangularwindowsdifferentphysicalchannels}
    \end{figure}

    Fig. \ref{fig:chapter02:continuousesddrectangularwindowsdifferentphysicalchannels} shows the influence of the parameters of the physical channel path on the spread onto the DD grid, assuming that the rectangular windows are employed. It is seen that, with rectangular windows, on-grid delay or Doppler shifts of the physical channel minimize the spread, while off-grid cases can result in significantly broader channel spread. This shows that the instantaneous values of the delay and Doppler shifts of the physical channel path can significantly affect the shape and sparsity level of the on-grid equivalence. Such a characteristic may be considered while using ESDD models for analyzing and designing systems operating over general physical channels.  
    
\section{Conclusions}
This paper revisits the on-grid delay-Doppler channel models which were studied by Bello in 1963. We extend  the results to account for more general windows applied to the transmitted or received signal in the time or frequency domain. It is shown that the cardinality and correlated channel gains of the on-grid channel model must be carefully chosen according to the transceiver and the instantaneous realization of the physical channel. Furthermore, rectangular time-frequency windows should be used in conjunction with the model. We hope the discussion can be useful for analyzing emerging waveforms targeting applications over doubly selective channels.

\section*{Acknowledgment}
The author wishes to thank Prof. Jinhong Yuan, Prof. Hai Lin, and Dr Akram Shafie for helpful discussions and suggestions, and thank the anonymous reviewers for their helpful comments. 

\bibliographystyle{IEEEtran}
\bibliography{ref}
\end{document}